%% file: main.tex
\documentclass[reprint,amsmath,amssymb,aps,prb,superscriptaddress]{revtex4-2}

\usepackage{graphicx}% Include figure files
\usepackage{dcolumn}% Align table columns on decimal point
\usepackage{bm}% bold math
\usepackage{bbold}
\usepackage[pdfusetitle,
            bookmarks=true,
            colorlinks,
            linkcolor=blue,
            citecolor=blue,
            urlcolor=blue
            ]{hyperref}
\usepackage{xcolor}
\usepackage{gensymb}
\usepackage{listings} 

\usepackage{physics}

\definecolor{codegreen}{rgb}{0,0.6,0}
\definecolor{codegray}{rgb}{0.5,0.5,0.5}
\definecolor{codepurple}{rgb}{0.58,0,0.82}
\definecolor{backcolour}{rgb}{0.95,0.95,0.92}

\lstdefinestyle{mystyle}{
    backgroundcolor=\color{backcolour},   
    commentstyle=\color{codegreen},
    keywordstyle=\color{magenta},
    numberstyle=\tiny\color{codegray},
    stringstyle=\color{codepurple},
    basicstyle=\ttfamily\footnotesize,
    breakatwhitespace=false,         
    breaklines=true,                 
    captionpos=b,                    
    keepspaces=true,                 
    numbers=left,                    
    numbersep=5pt,                  
    showspaces=false,                
    showstringspaces=false,
    showtabs=false,                  
    tabsize=2
}

\input{newcommands.tex}
\begin{document}

\title{Single-particle spectral function of fractional quantum anomalous Hall states}

% affiliation
\newcommand{\TUM}{\affiliation{Technical University of Munich, TUM School of Natural Sciences, Physics Department, 85748 Garching, Germany}}
\newcommand{\MCQST}{\affiliation{Munich Center for Quantum Science and Technology (MCQST), Schellingstr. 4, 80799 M{\"u}nchen, Germany}}
\newcommand{\Harvard}
{\affiliation{Department of Physics, Harvard University, Cambridge, Massachusetts 02138, USA}}

\author{Fabian Pichler} \TUM \MCQST
\author{Wilhelm Kadow} \TUM \MCQST
\author{Clemens Kuhlenkamp} \Harvard \TUM \MCQST
\author{Michael Knap} \TUM \MCQST

\date{\today}

\begin{abstract}
Fractional quantum Hall states are the most prominent example of states with topological order, hosting excitations with fractionalized charge. Recent experiments in twisted $\text{MoTe}_2$ and graphene-based heterostructures provided evidence of fractional quantum anomalous Hall (FQAH) states, which spontaneously break time-reversal symmetry and persist even without an external magnetic field. Understanding the unique properties of these states requires the characterization of their low-energy excitations. To that end, we construct a parton theory for the energy and momentum-resolved single-particle spectral function of FQAH states. We explicitly consider several experimentally observed filling fractions as well as a composite Fermi liquid in the half-filled Chern band. Charge fractionalization manifests itself in nearly momentum-independent spectra with a characteristic series of peaks determined from the filling fraction. The parton description qualitatively captures our numerical exact diagonalization results. Additionally, we discuss how the finite bandwidth of the Chern band and the non-ideal quantum geometry affect the fractionalized excitations.
Our work demonstrates that the energy and momentum-resolved electronic single-particle spectral function provides a valuable tool to characterize fractionalized excitations of FQAH states in moiré lattices.
\end{abstract}

\maketitle

\section{Introduction}
The low-energy excitations of strongly correlated phases of matter can exhibit remarkable properties such as charge fractionalization and anyonic exchange statistics. Fractional quantum Hall (FQH) states serve as an archetypical example of such phases with topological order, featuring low-energy excitations that are anyonic quasiparticles and quasiholes, each carrying a fraction of the electron charge~\cite{laughlinFQH1983}. Conventional FQH states are realized only under strong magnetic fields~\cite{Tsui1982}. Remarkably, fractional quantum anomalous Hall (FQAH) states, also known as fractional Chern insulators, were recently observed in twisted $\text{MoTe}_2$~\cite{cai_SignaturesFractional_2023, zeng_ThermodynamicEvidence_2023, park_ObservationFractionally_2023, xu_ObservationInteger_2023} and graphene heterostructures~\cite{xie_FractionalChern_2021, lu_FractionalQuantum_2024, aronsonDisplacementFieldcontrolledFractional2024}. Unlike FQH states, these FQAH states emerge even in the absence of an external magnetic field in fractionally filled bands with a topologically non-trivial Chern number. 

Experimentally, characterizing quantum anomalous Hall states currently relies on electronic compressibility measurements and magneto-optical spectroscopy techniques~\cite{cai_SignaturesFractional_2023, zeng_ThermodynamicEvidence_2023, foutty_MappingTwisttuned_2023}. Direct transport measurements have also been performed~\cite{li_QuantumAnomalous_2021, park_ObservationFractionally_2023, xu_ObservationInteger_2023}, but they remain challenging in transition metal dichalcogenide (TMD) heterostructures.
The development of the quantum twisting microscope~\cite{Inbar_QuantumTwisting_2023, birkbeckMeasuringPhononDispersion2024} introduced a new experimental tool for layered van der Waals materials that provides direct access to the energy and momentum-resolved spectra of both single-particle and collective excitations through tunneling spectroscopy~\cite{Peri_ProbingQuantum_2024, pichlerProbingMagnetismMoire2024, xiaoTheoryPhononSpectroscopy2024}.
Spectroscopic probes are valuable for advancing our understanding of the low-energy excitations in ``conventional'' FQH states, whose local density of states was demonstrated to show distinctive features of fractionalization~\cite{hu_HighResolutionTunneling_2024, gattuScanningTunnelingMicroscopy2024, pu_FingerprintsComposite_2023}. In the continuum the single-particle excitations do not carry any nontrivial momentum dependence. Understanding the role of the moiré band structure on the single-particle excitations in FQAH states remains an open problem. 

In this work, we develop a parton theory to compute the energy and momentum-resolved single-particle spectral function of such FQAH states, in which electrons (or holes) fractionalize into partons, each carrying a fraction of the electron charge. Considering the hole filling fractions of $\nu \in \{-1/3, -1/2, -2/3\}$, we compare the results from our parton description with exact diagonalization and find qualitative agreement; see Fig.~\ref{fig:1_v2} for an overview. As recent experiments have focused on hole doping, we will adopt that notion here and describe the parton theory for fractionalization of holes. Similar results hold for fractionalized electronic states. 
We demonstrate that the spectral function contains distinct signatures of charge fractionalization and the underlying moiré lattice structure. Charge fractionalization manifests in a nearly momentum-independent spectral function with a characteristic series of peaks uniquely determined by the filling fraction. The residual weak momentum dependence arises from interference effects associated with tunneling into the moir\'e lattice.
%In particular, we discuss the effects of the finite bandwidth and non-ideal quantum geometry of the Chern band on the spectrum of the fractionalized excitations, highlighting key differences to conventional FQH states.
Furthermore, we discuss the effects of the finite bandwidth and non-ideal quantum geometry of the Chern band on the spectrum of the fractionalized excitations, highlighting key differences from conventional FQH states.
Our findings demonstrate that momentum-resolved tunneling spectroscopy is a valuable tool to characterize the low-energy excitations of FQAH states and to improve our understanding of these states.

\section{Model} 
\begin{figure*}
\begin{center}
\includegraphics[width=0.99\linewidth]{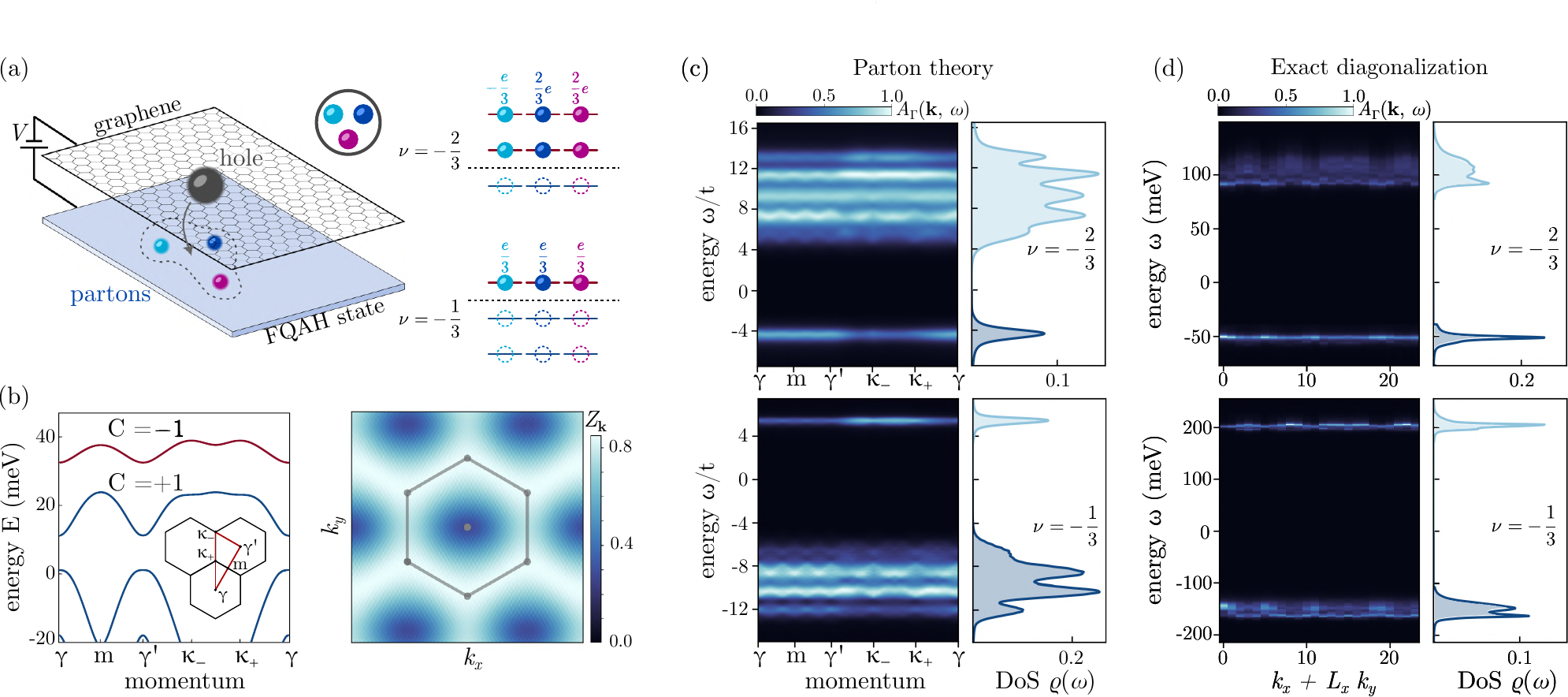}
\caption{\textbf{Single-particle spectral function of fractional quantum anomalous Hall states.} (a) Charge fractionalization for FQAH states. Left: In a tunneling experiment, a full hole (or electron) tunnels from a graphene layer (top layer) into an FQAH state (bottom layer), where it fractionalizes into partons. Right: Fractionalization pattern and schematic parton-band filling of $\nu = -2/3$ and $\nu = -1/3$ are illustrated. (b) Left: Bands for $\text{tMoTe}_2$ at a twist angle of $\theta = 3.7^\circ$. The two topmost bands have Chern numbers $C = -1$ and $C=+1$ in the $K$-valley. Right: Spectral weight distribution $Z_\kk$ in the Brillouin zone due to lattice interference effects for the topmost Chern band. Comparison of the single-particle spectral function $\mathcal{A}_\Gamma(\kk, \omega)$ computed using (c) a parton theory and (d) exact diagonalization. We focus here on $\mathcal{A}_\Gamma(\kk, \omega)$ instead of the experimentally measured spectral function $\mathcal{A}(\kk, \omega)$, which includes the geometric factor $Z_\kk$, to be able to clearly distinguish effects of non-ideal quantum geometry from bandstructure and lattice interference effects.}
\label{fig:1_v2}
\end{center}
\end{figure*}

Fractional quantum anomalous Hall states are strongly correlated states of matter, which form in a fractionally filled, topologically non-trivial band due to strong interactions relative to the bandwidth. They were first found numerically in lattice models~\cite{Sheng_FractionalQuantum_2011, Neupert_FractionalQuantum_2011, regnault_FractionalChern_2011} and more recently in continuum moir\'e Hamiltonians, describing twisted TMD homobilayers~\cite{Wu_TopologicalInsulators_2019, li_SpontaneousFractional_2021a, crepel_AnomalousHall_2023, reddy_FractionalQuantum_2023, xuMaximallyLocalizedWannier2024, wang_FractionalChern_2024, yu_FractionalChern_2024}. We focus on the moir\'e Hamiltonian for twisted $\text{MoTe}_2$, which for the $K$ valley is given by~\cite{Wu_TopologicalInsulators_2019}
\begin{equation}
    \mathcal{H} = \begin{pmatrix}
        - \frac{(\kk - \boldsymbol{\kappa}_+)^2}{2m^*}+\Delta_\bfr(\mathbf{r}) & \Delta_T(\mathbf{r}) \\
        \Delta^\dagger_T(\mathbf{r}) & - \frac{(\kk - \boldsymbol{\kappa}_-)^2}{2m^*}+\Delta_\tfr(\mathbf{r})
    \end{pmatrix} \label{eq:cont_2}
\end{equation}
with a layer $\ell \in \{\tfr=-1, \bfr=+1\}$ dependent moir\'e potential 
\begin{equation}
    \Delta_\ell(\mathbf{r}) =  \displaystyle\sum_{n \in \{0, 2, 4\}} 2 V\cos( \mathbf{G}_n \cdot \mathbf{r} + \ell \psi) \label{eq:moirepot_homo}
\end{equation}
and interlayer hopping term
\begin{equation}
    \Delta_T(\mathbf{r}) = w(1+e^{-i \mathbf{G}_1 \cdot \mathbf{r}}+e^{-i \mathbf{G}_2 \cdot \mathbf{r}})
\end{equation}
whose form is a consequence of symmetry considerations of the lattice~\cite{Bistritzer_MoireBands_2011, Wu_TopologicalInsulators_2019}. Here $\mathbf{G}_n = \mathrm{R}_{n\pi/3} \mathbf{G}_0$ and $\mathbf{G}_0 = 4\pi/\sqrt{3} a_M (1, 0)$ are moir\'e reciprocal lattice vectors, with the moir\'e lattice constant $a_M = a/\theta$. The lattice constant of $\text{MoTe}_2$ is $a=0.35$ nm. We adapt the model parameters $(w, V, \psi, m^*)$ from Ref.~\cite{wang_FractionalChern_2024}, which have been shown to reasonably describe the system near the experimentally relevant twist angle $\theta \sim 3.7^\circ$~\cite{yu_FractionalChern_2024}, where FQAH states have been observed~\cite{cai_SignaturesFractional_2023, zeng_ThermodynamicEvidence_2023, park_ObservationFractionally_2023, xu_ObservationInteger_2023}. In twisted $\text{MoTe}_2$ only the $\nu=-2/3$ state has been seen experimentally. Nevertheless, we also discuss the $\nu=-1/3$ state, which has been realized for other material platforms, such as graphene-based moir\'e lattices~\cite{xie_FractionalChern_2021, aronsonDisplacementFieldcontrolledFractional2024}. Since we are interested in only the spectral function within FQAH phases, we assume full spin-valley polarization, which can always be achieved experimentally by applying a small magnetic field and ramping it back to zero. Using Bloch's theorem, we write the single-particle Hamiltonian Eq.~\eqref{eq:cont_2}  in its eigenbasis as
\begin{equation}
    H_0 = \displaystyle\sum_\kk \displaystyle\sum_{\lambda} \eps_\kk^\lambda \gamma_{\lambda \kk}^\dagger \gamma_{\lambda\kk}, \label{eq:HamLat}
\end{equation}
where the creation operator for band $\lambda$ is given by
\begin{equation}
    \gamma^\dagger_{\lambda \kk} = \displaystyle\sum_{\mathbf{G}, \ell} \mathcal{U}^{\lambda}_{\mathbf{G}, \ell}(\kk) c^\dagger_{\ell, \kk + \mathbf{G}}, \label{eq:gammaOperator}
\end{equation}
summing over all moir\'e reciprocal lattice vectors $\mathbf{G}$ and layers $\ell$. In our notation, the operators $c$ and $\gamma$ annihilate an electron and thus, equivalently, can be interpreted to create a hole excitation. 
The first few bands $\eps_\kk^\lambda$ for a twist angle $\theta = 3.7^\circ$ are shown in Fig.~\ref{fig:1_v2}(b). The $\gamma$ operators effectively live on a triangular lattice, where each lattice site corresponds to an orbital superposition of the two layers and different reciprocal lattice vectors, encoded in the eigenvectors $\mathcal{U}_{\mathbf{G}, \ell}$. 
To capture the strongly-correlated FQAH states, interactions are essential. We introduce screened Coulomb interactions and project onto the topmost hole band of the Hamiltonian Eq.~\eqref{eq:HamLat}. The projected band has Chern number $C=-1$ and is comparatively flat at the experimentally relevant twist angle of $\theta \sim 3.7^\circ$.

\section{Spectral function of fractionalized states} \label{sec:SPSF}

The single-particle spectral function describes excitations above the ground state. The recent development of the quantum twisting microscope as a new experimental tool offers the possibility to directly access the energy and momentum-resolved spectral function through tunneling spectroscopy~\cite{Inbar_QuantumTwisting_2023}. In a tunneling process, one can add or remove only a full particle from the ground state, measuring the imaginary part of the retarded Green's function
\begin{equation}
    \mathcal{G}(\kk, \omega) = - i \displaystyle\int_0^\infty \dd t\; e^{i \omega t} \displaystyle\sum_{\mathbf{G}, \ell, \ell'} \langle \{ c_{\ell, \kk+\mathbf{G}}(t), c^\dagger_{\ell', \kk+\mathbf{G}}(0) \} \rangle. \label{eq:retGF1}
\end{equation}
After projecting to the topmost Chern band we have
\begin{equation}
    \mathcal{G}(\kk, \omega) = Z_\kk \Gamma(\kk, \omega)
    \label{eq:gfZ}
\end{equation}
with 
\begin{equation}
    Z_\kk = \displaystyle\sum_{\mathbf{G}, \ell, \ell'} \mathcal{U}_{\mathbf{G}, \ell}(\kk) \mathcal{U}^*_{\mathbf{G}, \ell'}(\kk) \label{eq:Zfac}
\end{equation}
determined by the bandstructure, and 
\begin{equation}
    \Gamma(\kk, \omega) = - i \displaystyle\int_0^\infty \dd t\; e^{i \omega t} \langle \{ \gamma_\kk(t), \gamma^\dagger_\kk(0) \} \rangle. \label{eq:GammaGF}
\end{equation}

We develop a parton theory for calculating the single-particle spectral function of states with fractionalized excitations and apply it specifically to FQAH states of the Jain sequence~\cite{jain_CompositefermionApproach_1989} at a filling of 
 \begin{equation}
 	\nu = -\frac{n}{2pn +1}
 \end{equation}
where $n$ is an integer and $p$ is a positive integer. 
Parton constructions for FQAH were established in Refs.~\cite{lu_SymmetryprotectedFractional_2012, mcgreevy_WaveFunctions_2012} and more recently in Ref.~\cite{lu_FractionalChern_2024}. In conventional FQH states, an analogous composite fermion approach has been used to calculate both the single-particle and collective responses~\cite{he_TunnelingTwodimensional_1993, Shytov_TunnelingEdgeCompressible1998, balram_VeryHighEnergyCollective_2022a, gattuScanningTunnelingMicroscopy2024, pu_FingerprintsComposite_2023, balram_SplittingGirvinMacDonaldPlatzman_2024, yue_ElectronicExcitations_2024}.

For Jain states, each hole fractionalizes into $2p+1$ partons, where $2p$ of them carry an electric charge $e^*= e n/(2pn +1)$ and fill bands with total Chern number $C=-1$, while one parton has electric charge $e^* = e/(2pn+1)$ and Chern number $C=-n$, Fig.~\ref{fig:1_v2}(a)~\cite{jainIncompressibleQuantumHall1989}. This results in the desired Hall conductivity of
 \begin{equation}
 	\sigma_{xy} = -\frac{n}{2pn +1} \frac{e^2}{2\pi \hbar}.
 \end{equation}
 Focusing on $p=1$, each hole in the projected Chern band fractionalizes into three partons 
 \begin{equation}
     \gamma(\mathbf{r}) = f_1(\mathbf{r}) f_2(\mathbf{r}) f_3(\mathbf{r}), \label{eq:threePartons}
 \end{equation}
introducing a gauge freedom, which can be used to couple the partons to an emergent gauge field. Through this gauge freedom, the individual parton states can break lattice symmetries $\mathrm{U}$ without breaking the translation symmetry of the underlying electronic state, as long as the parton state is invariant under a combination of $\mathrm{U}$ and some local gauge transformation $G_\mathrm{U}(\mathbf{r})$~\cite{Wen_QuantumOrdersSymmetric2002, lu_SymmetryprotectedFractional_2012}. In that way, it is possible to obtain gapped parton states, even at a fractional filling of the single-particle Chern band. The physical Hilbert space is obtained by enforcing the gauge constraint that the density of each parton has to equal the hole density. 

For concreteness, we focus on filling fractions of $\nu = - 1/3$ and $\nu = - 2/3$, corresponding to $n=1$ and $n=-2$, respectively. At a filling of $\nu = - 1/3$ there are three partons, each with a charge of $e^* = e/3$ and Chern number $C = -1$, while at a filling of $\nu = - 2/3$ there is one parton with charge $e^* = -e/3$ and Chern number $C =+2$, and two partons with charge $e^* = 2e/3$ and Chern number $C = -1$; see Fig.~\ref{fig:1_v2}(a).
We construct a parton mean-field ansatz for each parton species
\begin{equation}
    H_\text{parton} = \displaystyle\sum_{i, j}\displaystyle\sum_{\alpha=1}^{2p + 1} t^\alpha_{ij} f_\alpha^\dagger(\mathbf{r}_i)f_\alpha(\mathbf{r}_j),
\end{equation}
where the hopping terms $t^\alpha_{ij}$ should be understood to be self-generated through a mean-field decoupling from the other partons. We assume that no mixing terms between parton species are generated. To describe Jain states, an emergent gauge field that couples to the electric charge of the partons must be generated through the mean-field decoupling. We choose a gauge field
\begin{equation}
\mathbf{A}(\mathbf{r}) = \frac{4 \pi}{\sqrt{3} a^2 } \frac{\hbar}{e} (x + \sqrt{3} y ) \mathbf{\hat{e}_y} \label{eq:gaugeFieldFlux}
\end{equation}
generating a uniform flux, resulting in a Peierls phase of
\begin{equation}
t_{ij} \rightarrow t_{ij} \exp\left[{i \frac{e^*}{\hbar} \displaystyle\int_{\mathbf{x}_i}^{\mathbf{x}_j} \dd \mathbf{r} \cdot \mathbf{A}(\mathbf{r})}\right].
\end{equation}
In the limit of the integer quantum Hall effect, where $e^* = e$, the total number of flux quanta must equal the total number of electrons $N_\phi = N_e$, corresponding to one flux quantum per unit cell. This condition fixes the magnitude of the gauge field Eq.~\eqref{eq:gaugeFieldFlux}. Consequently, partons with charge $e^*$ feel a flux of $ 2\pi e^*/e$ per primitive unit cell of the lattice. Additional details and an explicit example for such a parton mean-field ansatz are given in Appendix~\ref{sec:appPartons}. 

On a mean-field level, the three parton flavors of Eq.~\eqref{eq:threePartons} are independent of each other. Therefore, the single-particle Green's function Eq.~\eqref{eq:GammaGF} in real space is the product of the three parton Green's functions:
\begin{equation}
    \Gamma(\mathbf{r}, \tau) = \mathcal{G}_{1}(\mathbf{r}, \tau)\mathcal{G}_{2}(\mathbf{r}, \tau) \mathcal{G}_{3}(\mathbf{r}, \tau) 
\end{equation}
In Fourier space, this yields the following convolution
\begin{align*}
    \Gamma(\kk, i \omega_n) = \int \dd \qb \dd \pbold &\displaystyle\sum_{\nu_m \nu_l} \mathcal{G}_1(\qb, i\nu_m)     \mathcal{G}_2(\pbold-\qb, i \nu_l - i\nu_m)  \\ &\times \mathcal{G}_3(\kk - \pbold, i \omega_n - i \nu_l) \numberthis \label{eq:parton_conv_GF}
\end{align*}
where $i \omega_n$ is a fermionic Matsubara frequency and $i \nu_{m/l}$ are bosonic Matsubara frequencies. The spectral function $\mathcal{A}_\Gamma(\kk, \omega) = - \Im \Gamma(\kk, \omega + i 0^+)/\pi$ of the $\gamma$ operator can be obtained by inserting the Lehman representation
\begin{equation}
    \mathcal{G}_\alpha(\kk, i \omega_n) = \displaystyle\int_{-\infty}^\infty \dd \eps\, \frac{\mathcal{A}_\alpha(\kk, \eps)}{i \omega_n - \eps}
\end{equation}
into Eq.~\eqref{eq:parton_conv_GF}. After calculating the Matsubara sums and performing the analytical continuation $i \omega_n \rightarrow \omega + i 0^+$, we take the imaginary part of the retarded Green's function to obtain the spectral function
\begin{align*}
    \mathcal{A}_\Gamma(\kk, \omega) = &\int \dd \qb \dd \pbold \int \dd \eps \dd \eps'\, \mathcal{A}_1(\qb,\eps) \mathcal{A}_2({\pbold-\qb}, \eps' - \eps) \\&\times \mathcal{A}_3(\kk - \pbold, \omega-\eps') [n_F(\eps) + n_B(\eps-\omega)]\\&\times[n_F(\eps'-\eps)-n_F(\eps'-\omega)],\numberthis \label{eq:parton_conv_SF_main}
\end{align*}
where $\mathcal{A}_\alpha(\kk, \omega)$ is the spectral function of the parton $f_\alpha$, $n_F(\eps)$ is the Fermi-Dirac distribution, and $n_B(\eps)$ is the Bose-Einstein distribution. 
The projected electronic spectral function is not simply the convolution of the parton spectral functions. The additional factors of Fermi-Dirac and Bose-Einstein distributions ensure that the particle and hole parts of the spectral function are not mixed. Physically, this can be understood from the fact that in a tunneling process, one always adds or removes a whole hole. Without the additional factors, there would be unphysical contributions to the total spectral function, where, for example, one parton is added, and two partons are removed.

We obtain the experimentally accessible single-particle spectral function from Eq.~\eqref{eq:gfZ} as
\begin{equation}
    \mathcal{A}(\kk, \omega) =  Z_\kk \mathcal{A}_\Gamma(\kk, \omega) 
    \label{eq:fullELSF}
\end{equation}
and compute it at fractional fillings $\nu = -2/3$ and $\nu=-1/3$;  Fig.~\ref{fig:1_v2}(c). To facilitate the comparison between our parton and exact diagonalization results, we show $\mathcal{A}_\Gamma(\kk, \omega)$ in Fig.~\ref{fig:1_v2}(c, d), instead of the experimentally accessible spectral function $\mathcal{A}(\kk, \omega)$, which includes the geometric factor $Z_\kk$ of Eq.~\eqref{eq:Zfac}. The full spectral function Eq.~\eqref{eq:fullELSF} is shown in Appendix~\ref{sec:EMSF} in Fig.~\ref{fig:fullSF}. For simplicity, we consider only nearest-neighbor hopping in our mean-field parton description. Including further-range hopping terms would modify the parton band structure and, consequently, the details of the spectral function, such as the precise value of the gap. However, since we do not expect our parton description to reliably capture such quantitative details anyway, we focus our discussion for now on qualitative behavior. 

The momentum convolution in Eq.~\eqref{eq:parton_conv_SF_main} smears out any momentum dependence of the parton bands, resulting in an effectively momentum-independent spectral function $\mathcal{A}_\Gamma(\kk, \omega)$ of the $\gamma$ particles. Physically, this is a signature of deconfined fractionalized particles.
By contrast, the experimentally-accessible electronic spectral function $\mathcal{A}(\kk, \omega)$ 
has a momentum dependence resulting from $Z_\kk$, see Fig.~\ref{fig:1_v2}(b) and Fig.~\ref{fig:fullSF} in Appendix~\ref{sec:EMSF}. This non-trivial momentum dependence introduced by $Z_\kk$ is thus a lattice interference effect between the orbitals, which occupy each lattice point of the projected band and does not depend on the underlying FQAH state. The spectral weight is minimal in a region around the $\Gamma$ point and maximal at the boundary of the Brillouin zone. 
For comparison, a fully layer-polarized, topologically trivial state, which can be obtained by applying a strong displacement field, has an approximately constant $Z_\kk$ since there is no interference between the two layers anymore.

Having analyzed the moiré bandstructure effects on the spectrum, we now turn to the energy dependence of the bands and focus our discussion on the total density of states $\rho(\omega) = \sum_\kk \mathcal{A}(\kk, \omega)/N$. We find results qualitatively similar to those of Refs.~\cite{gattuScanningTunnelingMicroscopy2024, pu_FingerprintsComposite_2023}, in which the local density of states for conventional FQH states is calculated, and Ref.~\cite{hu_HighResolutionTunneling_2024}, in which the local densities of states of various FQH states are measured. The spectral function shows a clear gap between the particle and hole excitations; a feature of the incompressible FQAH states. At filling $\nu=-2/3$, a single sharp peak is present at negative energies.
This can be understood as follows: For this filling fraction, two out of three parton bands are filled with holes; see Fig.~\ref{fig:1_v2}(a). 
Adding a hole thus requires adding three partons, one for each flavor, to the remaining empty band. As a result, the peak in the density of states is located at the sum of the energies of the three empty parton bands.
The response for positive energies is qualitatively different. It consists of a broad continuum with small peaks on top. Since there are several filled parton bands, the many combinations of removing three partons result in a much broader signal. 
The reverse applies for the state at a filling of $\nu = -1/3$, where the sharp peak is located at positive energies and the broader continuum at negative energies.

We find qualitative agreement between the parton results in Fig.~\ref{fig:1_v2}(c) and exact diagonalization results in Fig.~\ref{fig:1_v2}(d), supporting the parton ansatz to accurately capture the low-energy excitations of the FQAH states. Technical details regarding our exact diagonalization calculation are given in Appendix~\ref{sec:EDdetails}. The lack of dispersing bands and the characteristic sequence of peaks in the density of states are distinctive signatures of fractionalization. These properties distinguish FQAH states from competing states, such as Fermi liquids and charge density waves~\cite{pichlerProbingMagnetismMoire2024}.

\section{Effects of band dispersion and quantum geometry}

\begin{figure}
\begin{center}
    \includegraphics[width=\linewidth]{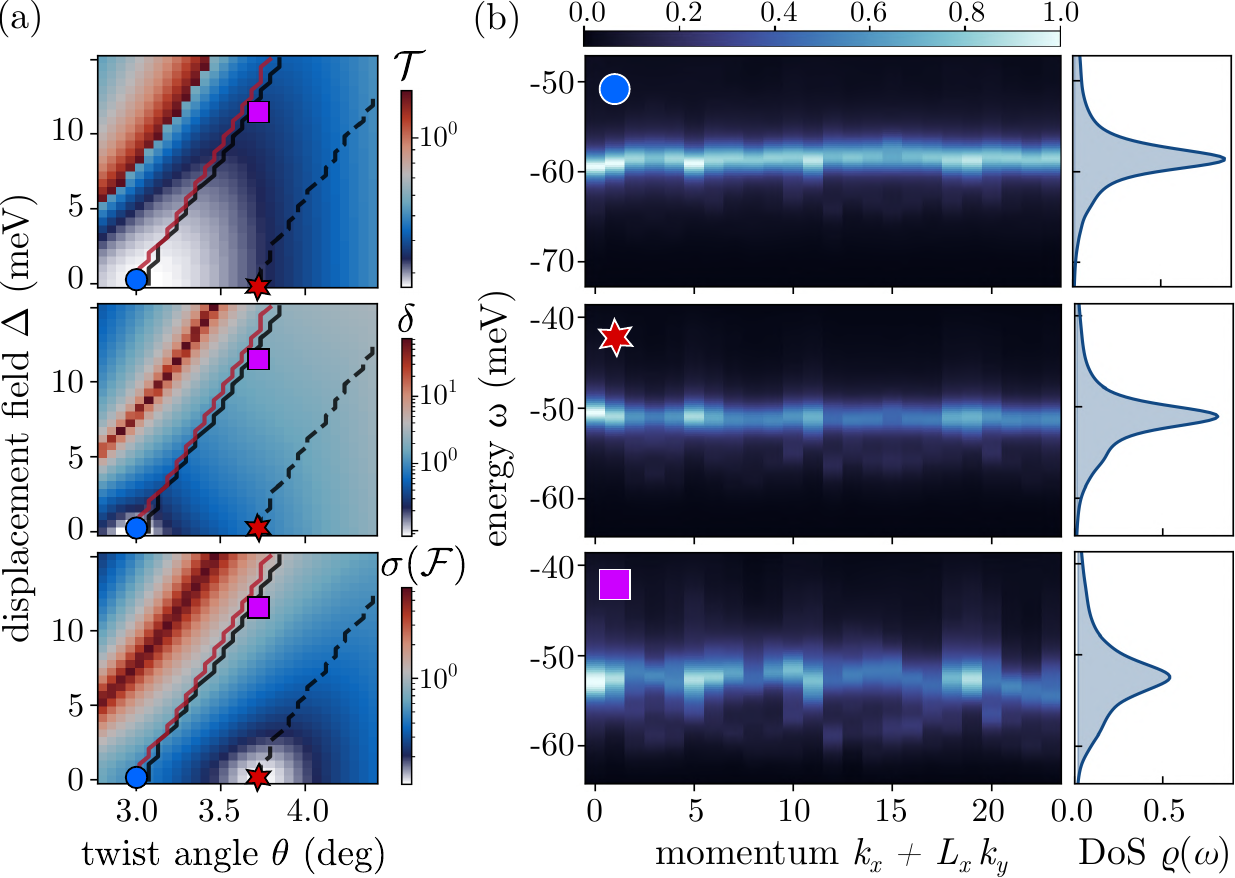}
    \caption{\textbf{Finite bandwidth and non-ideal quantum geometry.} (a) From top to bottom, violation of the trace condition $\mathcal{T}$, flatness ratio $\delta$, and Berry curvature fluctuations $\sigma(\mathcal{F})$ as a function of twist angle and displacement field. The red, solid black, and dashed black lines correspond to the minimum of $\mathcal{T}$, $\delta$, and $\sigma(\mathcal{F})$ for a fixed displacement field, respectively. A small flatness ratio $\delta$ strongly correlates with a weak violation of the trace condition $\mathcal{T}$. (b) Hole spectral function $\mathcal{A}_\Gamma(\kk, \omega)$ for three different values of the twist angle and displacement field at filling $\nu = -2/3$. A change in the twist angle has little effect on the hole spectral function, while a weak displacement field broadens the density of states and introduces small inhomogeneities in the momentum distribution.}
    \label{fig:Robustness}
\end{center}
\end{figure}

Conventional FQH states are realized in the lowest Landau level, which has a perfectly flat dispersion and is characterized by an ideal quantum geometry, i.e., it has a uniform Berry curvature distribution and fulfills the trace condition~\cite{royBandGeometryFractional2014}
\begin{equation}
    \mathcal{T} := \int \frac{\dd^2 k}{(2\pi)^2} \big[ \Tr g(\kk) - \mathcal{F}(\kk) \big] = 0 \label{eq:tracecond}
\end{equation}
where $g(\kk)$ is the quantum metric and $\mathcal{F}(\kk)$ is the Berry curvature, being the real and imaginary part of the quantum geometric tensor, respectively~\cite{provostRiemannianStructureManifolds1980, restaInsulatingStateMatter2011}. We calculate them using~\cite{panatiTrivialityBlochBloch2007, brouder_ExponentialLocalizationWannier2007} 
\begin{subequations}
\begin{align}
    g(\kk) &= \frac{1}{2} \Tr\left( \frac{\partial P(\kk)}{\partial k_i} \frac{\partial P(\kk)}{\partial k_i} \right), \\
    \mathcal{F}(\kk) &= \frac{i}{2} \eps_{i j} \Tr \left( P(\kk) \left[\frac{\partial P(\kk)}{\partial k_i}, \frac{\partial P(\kk)}{\partial k_j}\right] \right),
\end{align}
\end{subequations}
where $P(\kk) = \mathcal{U}(\kk) \mathcal{U}^\dagger(\kk)$ is the projector to the band with eigenvector $\mathcal{U}(\kk)$; see Eq.~\eqref{eq:gammaOperator}.
The flat dispersion, uniform Berry curvature, and the trace condition Eq.~\eqref{eq:tracecond} of the projected lowest Landau level result in a particle-hole symmetry between states at a filling of $\nu$ and $1-\nu$. This exact particle-hole symmetry is explicitly broken in the system we consider because the Chern band has a finite dispersion, and its quantum geometry is non-ideal. Consequently, the spectral functions of the $\nu = -1/3$ and $\nu=-2/3$ states are not particle-hole symmetric with respect to each other. Nevertheless, an approximate particle-hole symmetry on a qualitative level is retained, as shown in Fig.~\ref{fig:1_v2}(c, d). Another consequence of the finite dispersion on the lattice is that the peaks in the density of states are intrinsically broadened compared to the sharp peaks expected for flat Landau levels. 

Next, we discuss the influence of a finite band dispersion and a non-ideal quantum geometry on the hole spectral function of FQAH states. Both the band dispersion and the quantum geometry are single-particle properties, which can be tuned by changing the twist angle $\theta$ and applying a displacement field $\Delta$, which acts as a sublattice-staggered chemical potential. To characterize the quantum geometry, we use the violation of the trace condition $\mathcal{T}$ of Eq.~\eqref{eq:tracecond}, the Berry curvature fluctuations $\sigma(\mathcal{F})$, which we define as the standard deviation of the Berry curvature, and the flatness ratio $\delta$, which is the ratio between the bandwidth of the topmost band and the gap between the two topmost bands.

We compute $\mathcal{T}$, $\sigma(\mathcal{F})$, and $\delta$ as a function of the twist angle and an applied displacement field; Fig.~\ref{fig:Robustness}(a). The closing of the single-particle gap can be observed for large displacement fields via the diverging flatness ratio and Berry curvature fluctuations. At zero displacement field, the minimum in both the flatness ratio and trace condition violation is achieved at a twist angle of $\theta \sim 3.0^\circ$. By contrast, the Berry curvature fluctuations are minimal for $\theta \sim 3.7^\circ$, where the FQAH states have been observed in experiments~\cite{cai_SignaturesFractional_2023, zeng_ThermodynamicEvidence_2023, park_ObservationFractionally_2023, xu_ObservationInteger_2023}. In Fig.~\ref{fig:Robustness}(b), we show the hole part of the single-particle spectral function $\mathcal{A}_\Gamma(\kk, \omega)$ at $\nu = -2/3$ for three different values of the twist-angle and displacement field, representative of three different parameter ranges of the flatness ratio and quantum geometry. We show $\mathcal{A}_\Gamma(\kk, \omega)$ again to be able to clearly distinguish non-ideal quantum geometry from bandstructure effects. For an ideal FQH state, we expect a $\delta$-like peak with no momentum dependence for the hole spectrum. 
Determining how close the spectral function of an FQAH state is to this ideal behavior can provide insight into how closely the state approximates an ideal FQH state. 

At zero displacement field, we do not observe a strong dependence of the spectral function on the twist angle. The flatter band with weaker violation of the trace condition at $\theta=3.0^\circ$ shows a slightly higher uniformity in momentum space, compared to the state with nearly uniform Berry curvature at $\theta=3.7^\circ$. This suggests that a flat band with a good trace condition is more crucial than uniform Berry curvature, agreeing with recent studies~\cite{ledwithVortexabilityUnifyingCriterion2023, morales-duranPressureenhancedFractionalChern2023, andrewsStabilityFractionalChern2024}.
When a small displacement field is applied, such that the system still remains within the FQAH phase, more pronounced differences emerge. The spectral function reveals a broadened peak that is less uniform in momentum space, indicating that the quality of the FQAH states, as measured by their proximity to an ideal FQH state in the lowest Landau level, deteriorates, as expected. 
This shows how measuring the single-particle spectral function in an experiment can be used to determine the quality of FQAH states and how close they are to a competing state, such as a charge density wave. 

We find that the size of the energy gap in FQAH states only weakly depends on single-particle properties such as band dispersion and quantum geometry. Instead, the gap in the spectral function is essentially determined by the interaction strength. This emphasizes the dominant role of electron-electron interactions over single-particle characteristics in determining the single-particle gap of FQAH states.

\section{Composite Fermi Liquid} 

\begin{figure}
\begin{center}
\includegraphics[width=\linewidth]{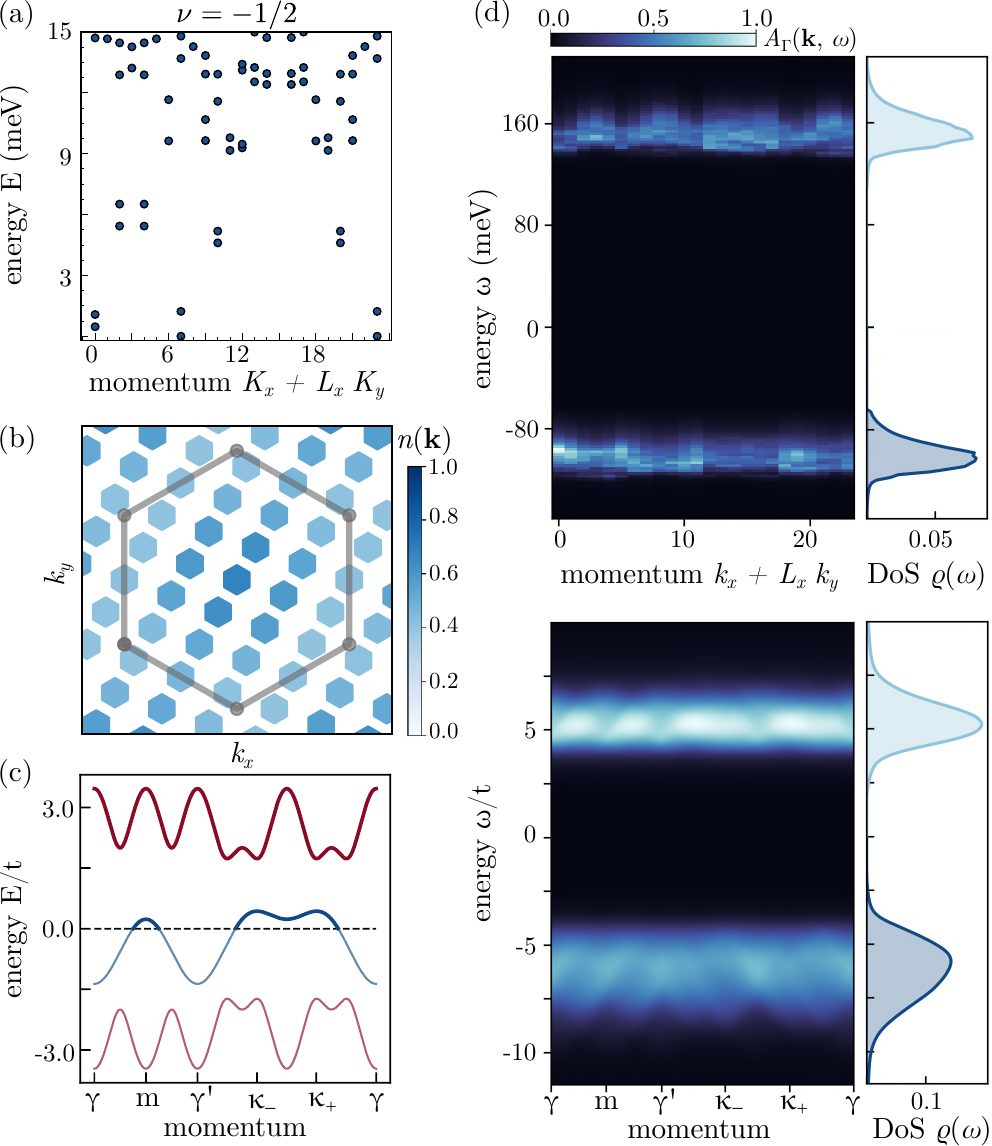}
\caption{\textbf{Composite Fermi liquid in the half-filled Chern band.} (a) Eigenvalues for the half-filled Chern band at $\theta = 3.7^\circ$ for $L_x \times L_y = 6 \times 4$ sites, consistent with the energy spectrum of the lowest Landau level. (b) Nearly uniform particle density $n_\kk$ of the composite Fermi liquid. (c) Parton bands for the Fermi sea (blue) and the Laughlin $\nu=-1/2$ state (red). The Fermi sea breaks particle-hole symmetry. (d) Comparison of the single-particle spectral function of a composite Fermi liquid, obtained from exact diagonalization (top) and parton theory (bottom).}
\label{fig:3_v1}
\end{center}
\end{figure}

So far, we have focused on incompressible FQAH states of the Jain sequence. Next, we consider the half-filled topmost Chern band of twisted $\text{MoTe}_2$, which, in analogy to the compressible composite Fermi liquid (CFL) of the half-filled lowest Landau level~\cite{Halperin_TheoryHalffilled_1993, wang_ParticleHoleSymmetry_2017}, is expected to host an anomalous CFL~\cite{dong_CompositeFermi_2023, goldmanZeroFieldComposite2023, reddyTowardGlobalPhase2023}. Experimental signatures of such a topological non-trivial state have been observed in transport measurements, while no optical gap was observed~\cite{park_ObservationFractionally_2023, cai_SignaturesFractional_2023}, consistent with a compressible state. In Fig.~\ref{fig:3_v1}(a), we show the energy eigenvalues of the half-filled topmost Chern band. The energy spectrum is consistent with the corresponding spectrum of the half-filled Landau level, agreeing with Refs.~\cite{dong_CompositeFermi_2023, goldmanZeroFieldComposite2023, reddyTowardGlobalPhase2023}. A distinctive signature of the CFL is a uniform particle density $n_\kk$ across the Brillouin zone~\cite{dong_CompositeFermi_2023, reddyTowardGlobalPhase2023}, which is in agreement with our numerical results; Fig.~\ref{fig:3_v1}(b).

We describe the half-filled Chern band that hosts an anomalous CFL with a parton ansatz; each hole fractionalizes into a charge-$e$ boson $b$ and chargeless fermion $f$~\cite{barkeshli_ContinuousTransitions_2012, barkeshli_ParticleholeSymmetry_2015}:
\begin{equation}
    \gamma = b f. \label{eq:partonCFL}
\end{equation}
The boson is assumed to be in a $\nu = -1/2$ Laughlin state, while the fermion fills up a topologically trivial Fermi sea. The resulting electronic state is compressible, but its single-particle spectral function is gapped. This can be intuitively understood as follows. The two-point correlator $\langle \gamma^\dagger_i \gamma_j \rangle \sim \langle b^\dagger_i b_j \rangle  \langle f^\dagger_i f_j \rangle$ results in a gapped spectral function, due to the exponentially decaying correlations in $\langle b^\dagger_i b_j \rangle$ of the gapped $\nu=-1/2$ Laughlin state. By contrast, the density response will be essentially the same as for a Fermi liquid, since the $\nu = -1/2$ Laughlin state has uniform density: $\langle \gamma^\dagger_i \gamma_i \gamma^\dagger_j \gamma_j \rangle \sim \langle b^\dagger_i b_i b^\dagger_j b_j \rangle \langle f^\dagger_i f_i f^\dagger_j f_j\rangle \sim n_b^2 \langle f^\dagger_i f_i f^\dagger_j f_j\rangle$. The intuitive picture agrees well with numerical results for the dynamical density response of CFLs~\cite{Kumar_NeutralExcitations_2022}. These properties render the detection of CFLs using conventional magneto-optical techniques challenging, as they do not observe any gap, which can be, however, directly measured with tunneling spectroscopy.

For the boson $b$, we use a second-level parton construction, assuming that the boson fractionalizes into two fermions, each carrying charge $e^* = e/2$ and filling up a Chern band with $C=-1$. The gapped parton bands are again obtained by coupling the partons to the gauge field Eq.~\eqref{eq:gaugeFieldFlux}. The chargeless fermion $f$ in Eq.~\eqref{eq:partonCFL} forms a Fermi sea with a nearly circular Fermi surface. Concretely, the single band is obtained from a nearest-neighbor tight-binding Hamiltonian on a triangular lattice at half-filling. The parton band structure for the different parton species is shown in Fig.~\ref{fig:3_v1}(c). We compute the resulting spectral function using Eqs.~\eqref{eq:parton_conv_SF_main} and \eqref{eq:fullELSF}. Due to the convolution with the Fermi sea, there are no sharp peaks but rather a broad continuum. The particle and hole sides of the spectral function are separated by a gap stemming from the $\nu = - 1/2$ Laughlin state. These results are consistent with previous theoretical predictions for the CFL in the half-filled Landau level~\cite{he_TunnelingTwodimensional_1993, kim_InstantonsSpectral_1994, yue_ElectronicExcitations_2024}. 
A major difference from the half-filled Landau level is the explicit breaking of particle-hole symmetry due to a finite dispersion and non-uniform Berry curvature in the half-filled Chern band that we consider; (cf. Ref.~\cite{reddyTowardGlobalPhase2023}). This particle-hole asymmetry also manifests in the spectral function, where the particle side shows a sharper drop-off for high energies. On a parton level, the particle-hole asymmetry is introduced through the dispersion of the trivial fermion $f$.

We remark that particle-hole asymmetry was experimentally observed in the lowest half-filled Landau level of graphene~\cite{hu_HighResolutionTunneling_2024}. There the origin of the observed asymmetry may be an artifact of the measurement process, but the precise reason is still an open question.

\section{Conclusions and Outlook} \label{sec:conclusion}
Our results show that FQAH states leave characteristic signatures in the energy and momentum-resolved single-particle spectral function, which is experimentally accessible in van der Waals heterostructures via the quantum twisting microscope~\cite{Inbar_QuantumTwisting_2023}. Our parton theory qualitatively captures numerical simulations and thus provides a clear interpretation of the observed spectral features. We focused on a regime where contributions from higher bands are far off in energy. For future work, it will be interesting to explore the effects of band mixing when this assumption is no longer fulfilled. Furthermore, one could explore the collective response of FQAH states, which can be measured with quantum twisting microscopes by inelastic tunneling~\cite{Peri_ProbingQuantum_2024, pichlerProbingMagnetismMoire2024, birkbeckMeasuringPhononDispersion2024, xiaoTheoryPhononSpectroscopy2024}. To obtain a more quantitatively accurate spectral function within a parton theory, it is essential to optimize the parton hopping parameters, for example, by variational approaches.

The single-particle spectral function was proposed to detect and characterize other states with topological order as well. For example, in quantum spin liquids, previous theoretical work already highlighted that signatures of fractionalized excitations can be observed in the spectral function~\cite{Lauchli2004, Senthil_TheoryContinuous_2008, Podolsky_MottTransitionSpin2009, Kadow_HoleSpectral_2022, kuhlenkamp_ChiralPseudospin_2024, Kadow2024, nyhegnProbingQuantumSpin2024}. Other, rather unexplored topological states include anomalous Hall crystals~\cite{pan_TopologicalPhases_2022, dongTheoryQuantumAnomalous2024, dong_AnomalousHall_2023, song_IntertwinedFractional_2023, sheng_QuantumAnomalous_2024, gonçalves2024dopinginducedquantumanomaloushall}, non-Abelian FQAH states~\cite{liu_NonAbelianFractional_2024, chen_RobustNonAbelian_2024, wang_HigherLandauLevel_2024}
and fractional topological insulators~\cite{Bernevig_QuantumSpinHall2006, LevinFTI2009, titusFTI2011, kwanWhenCouldAbelian2024}. The latter consist of two time-reversal symmetric copies of FQH states and hence show no Hall conductivity, rendering their experimental detection challenging. We expect the single-particle spectral function of fractional topological insulators to resemble that of FQH states, providing a possible detection scheme for these otherwise elusive states.

\ \\{\textbf{Acknowledgements}} \\
{
We acknowledge support from the Deutsche Forschungsgemeinschaft (DFG, German Research Foundation) under Germany’s Excellence Strategy--EXC--2111--390814868, TRR 360 – 492547816 and DFG grants No. KN1254/1-2, KN1254/2-1, the European Research Council (ERC) under the European Union’s Horizon 2020 research and innovation programme (grant agreement No. 851161), as well as the Munich Quantum Valley, which is supported by the Bavarian state government with funds from the Hightech Agenda Bayern Plus. C.K. acknowledges funding from the Swiss National Science Foundation (Postdoc.Mobility Grant No. 217884).
}

\ \\{\textbf{Data availability}} \\Data, data analysis, and simulation codes are available upon reasonable request on Zenodo~\cite{zenodo}.

\appendix

\section{Parton construction} \label{sec:appPartons}

\begin{figure}
\begin{center}
    \includegraphics[width=\linewidth]{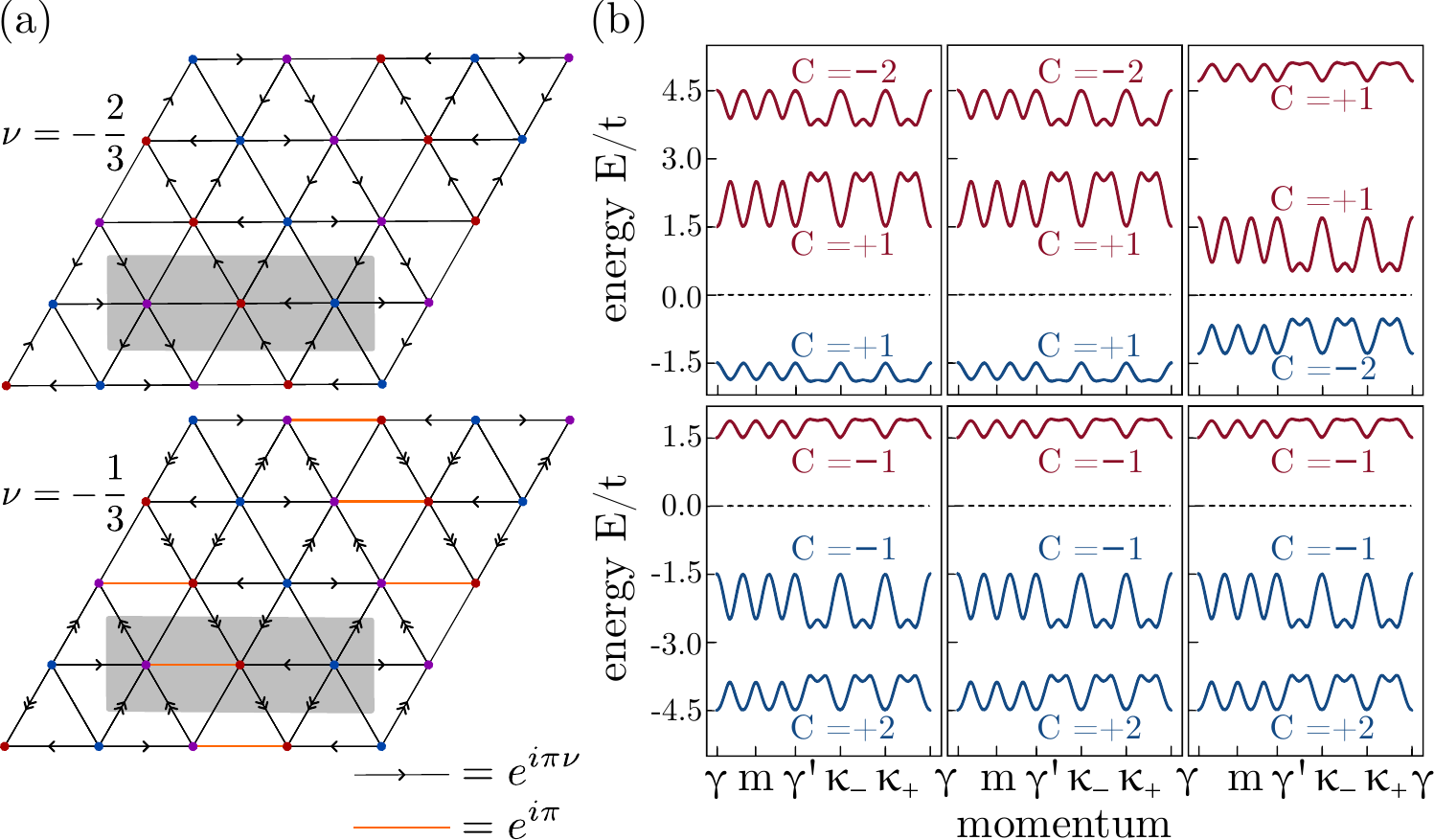}
    \caption{\textbf{Parton theory.} (a) Flux pattern for the parton mean-field ansatz with $4\pi/3$ (top) and $2\pi/3$ (bottom) flux per primitive unit cell. A particle hopping along a single black arrow picks up a phase of $e^{i\pi\nu}$. Two black arrows yield a phase of $e^{i 2\pi \nu}$, while an orange line adds a phase $e^{i \pi}$. The enlarged unit cell is highlighted in gray. (b) Parton bands. For $\nu=-2/3$ (top), two partons have charge $e^* = 2e/3$ and total Chern number $C=-1$, while the third parton has charge $e^* = -e/3$ and Chern number $C=+2$. For $\nu = -1/3$ (bottom), all three partons are identical and fill up the same bands, with total Chern number $C=-1$. }
    \label{fig:Partons}
\end{center}
\end{figure}

We describe the partons using a mean-field ansatz
\begin{equation}
    H_\text{parton} = \displaystyle\sum_{i, j}\displaystyle\sum_{\alpha=1}^{2p + 1} t^\alpha_{ij} f_\alpha^\dagger(\mathbf{r}_i)f_\alpha(\mathbf{r}_j)
\end{equation}
where we have assumed that different flavors of partons do not mix. Each parton species couples to the emergent gauge field Eq.~\eqref{eq:gaugeFieldFlux}, breaking lattice translation symmetry and resulting in an enlarged magnetic unit cell. As discussed in the main text, the underlying electronic state still respects all lattice symmetries $\mathrm{U}$, as long as the parton state is invariant under a combination of $\mathrm{U}$ and some local gauge transformation $G_\mathrm{U}(\mathbf{r})$.

Partons with charge $e^* = \pm e/3$ feel a flux of $\pm 2\pi/3$ per primitive unit cell, while partons with charge $e^* = 2e/3$ feel $4\pi/3$ flux per primitive unit cell, corresponding to $2\pi/3$ flux per triangle of the triangular lattice. For the gauge choice given by Eq.~\eqref{eq:gaugeFieldFlux}, the corresponding flux pattern for nearest-neighbor hopping is shown in Fig.~\ref{fig:Partons}(a). As an example, the $4\pi/3$-flux parton Hamiltonian can be written in the sublattice basis $\Psi_\kk = (f_{A \kk}, f_{B \kk}, f_{C \kk})^T$ as
\begin{equation}
    h_\kk^{(1)} = \begin{bmatrix} e^{-i \kk \cdot \mathbf{a}_2} & e^{-i \varphi} e^{-i \kk \cdot \mathbf{a}_1} & e^{i \varphi} e^{-i \kk \cdot \mathbf{a}_3} \\
    e^{-i \varphi}e^{-i \kk \cdot \mathbf{a}_3} & e^{i \varphi} e^{-i \kk \cdot \mathbf{a}_2} & e^{-i \kk \cdot \mathbf{a}_1} \\
    e^{i \varphi} e^{-i \kk \cdot \mathbf{a}_1} & e^{-i \kk \cdot \mathbf{a}_3} & e^{-i \varphi}e^{-i \kk \cdot \mathbf{a}_2}
    \end{bmatrix}
\end{equation}

\begin{equation}
    H_{4\pi/3} = - t \displaystyle\sum_\kk \Psi^\dagger_\kk \big( h_\kk^{(1)} + \text{h.c} \big) \Psi_\kk
\end{equation}
with $\varphi= 2\pi/3$. The basis lattice vectors are given by $\mathbf{a}_{1/2} = a_M (\pm \sqrt{3}, 1)/2$ and $\mathbf{a}_3 = - (\mathbf{a}_1 + \mathbf{a}_2)$. In Fig.~\ref{fig:Partons}(b), we show the parton bands and their corresponding Chern numbers. Note that we consider states with hole doping and, as a consequence, fill the parton bands from above with holes. For $\nu = -1/3$, only a single parton band is filled, with Chern number $C=-1$ for all three parton flavors. Conversely, for $\nu=-2/3$, two bands are filled for each parton species. The filled bands for the two partons with charge $e^*=2e/3$ have total Chern number $C=-1$, while the filled bands for the single parton with charge $e^*=-e/3$ have total Chern number $C=+2$. \\

For a generic Jain state at a filling of $\nu = - n /(2 p n + 1)$, each hole fractionalizes into $2p+1$ partons. Equation~\eqref{eq:parton_conv_SF_main} for the spectral function $\mathcal{A}_\Gamma(\kk, \omega)$ discussed in the main text for $p=1$ generalizes for general positive integers $p$ as follows:
\begin{align*}
    \mathcal{A}_\Gamma(\kk, \Omega) &= \int \displaystyle\prod_{\beta = 1}^{2p}\dd\qb_{\beta}\dd\eps_{\beta} \displaystyle\prod_{\alpha=1}^{2p+1} \mathcal{A}(\qb_\alpha -\qb_{\alpha-1}, \eps_\alpha -\eps_{\alpha-1}) \\
    &\times \displaystyle\prod_{j = 1}^{p} \bigg\{[n_F(\eps_{2j -1}-\eps_{2j-2}) + n_B(\eps_{2j-1}-\eps_m)]\\
    &\times[n_F(\eps_{2j}-\eps_{2j-1})-n_F(\eps_{2j}-\eps_m)]\bigg\}, \numberthis
\end{align*}
with $\qb_0 \equiv 0$, $\qb_{2p+1} \equiv \kk$, $\eps_0 \equiv 0$, and $\eps_{2p+1} \equiv \Omega$. 

\section{Experimentally accessible spectral function} \label{sec:EMSF}

\begin{figure}
\begin{center}
    \includegraphics[width=\linewidth]{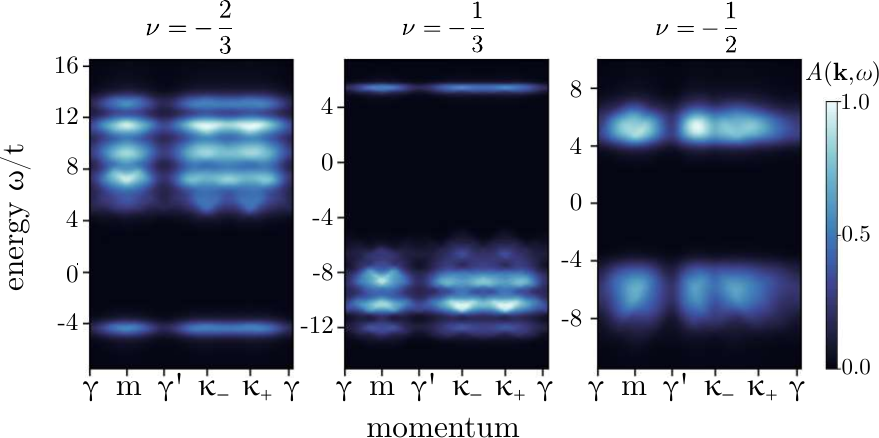}
    \caption{\textbf{Experimentally accessible single-particle spectral function.} Full spectral function $\mathcal{A}(\kk, \omega)$ from parton theory, including the geometric factor $Z_\kk$, which is due to interference effects between the two layers. The three fillings discussed in the main text, $\nu = -2/3$, $\nu=-1/3$, and $\nu=-1/2$, are shown from left to right.}
    \label{fig:fullSF}
\end{center}
\end{figure}
The experimentally accessible single-particle spectral function $\mathcal{A}(\kk, \omega)$, Eq.~\eqref{eq:fullELSF}, includes the geometric factor $Z_\kk$ in Eq.~\eqref{eq:Zfac}, which introduces a non-trivial momentum dependence that results from the bandstructure, as discussed in Sec.~\ref{sec:SPSF}. We show the full spectral function $\mathcal{A}(\kk, \omega)$ for the three filling factors $\nu \in \{-2/3, -1/3, -1/2\}$ in Fig.~\ref{fig:fullSF}.

\section{Details on exact diagonalization} \label{sec:EDdetails}

\begin{figure}
\begin{center}
    \includegraphics[width=\linewidth]{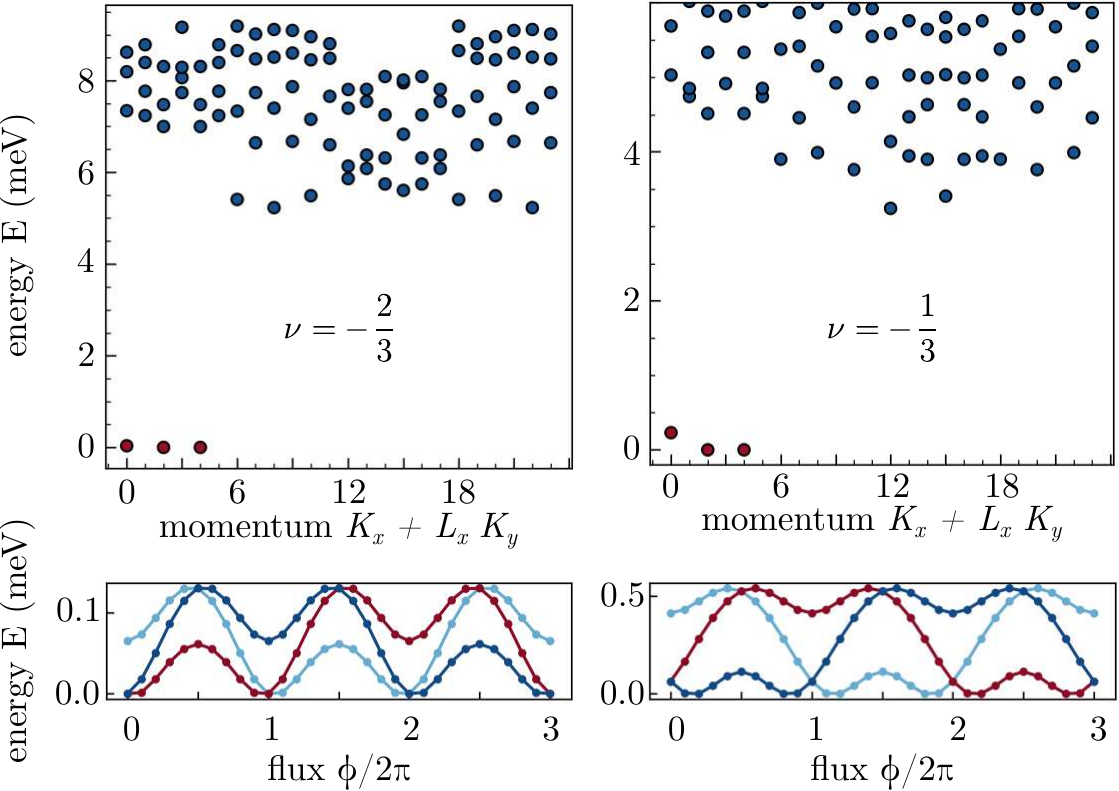}
    \caption{\textbf{Exact diagonalization energy eigenvalues.} The three degenerate ground states of FQAH states on a torus for $\nu = -1/3$ (right) and $\nu = -2/3$ (left) show a clear gap to the first excited states. Bottom: The ground states flow into each other under flux insertion. They only return to the original configuration after inserting three flux quanta.}
    \label{fig:ED}
\end{center}
\end{figure}

For our exact diagonalization calculations, we add double-gated screened Coulomb interactions
\begin{equation}
    V(\qb) = \frac{e^2}{4\pi \epsilon_0 a_M} \frac{4 \pi \xi}{\sqrt{3} \epsilon a_M} \frac{\text{tanh}(|\qb|\xi)}{|\qb |\xi} \label{eq:screenedColoumb}
\end{equation}
to the single-particle Hamiltonian Eq.~\eqref{eq:HamLat}. We use a screening length $\xi = a_M$ and a dielectric constant $\epsilon = 10$. After projecting to the topmost Chern band, the interaction Hamiltonian is given by
\begin{equation}
    \Bar{H}_\text{int} = \frac{1}{L_x L_y} \displaystyle\sum_{\qb \kk \kk'} V_{\kk \kk'}(\qb) \gamma^\dagger_{\kk + \qb} \gamma^\dagger_{\kk' -\qb} \gamma_{\kk'} \gamma_{\kk}
\end{equation}
with 
\begin{equation}
    V_{\kk \kk'}(\qb) = \displaystyle\sum_\mathbf{G} V(\qb + \mathbf{G}) \lambda_{\qb + \mathbf{G}}(\kk) \lambda_{-\qb - \mathbf{G}}(\kk'),
\end{equation}
where we defined the form factors
\begin{equation}
    \lambda_\qb(\kk) = \displaystyle\sum_{\mathbf{G}, \ell} \mathcal{U}^*_{\mathbf{G}, \ell}(\kk+\qb) \mathcal{U}_{\mathbf{G}, \ell}(\kk)
\end{equation}
using the eigenvectors $\mathcal{U_{\mathbf{G}, \ell}}(\kk)$ of the Chern band, see Eq.~\eqref{eq:gammaOperator}. The total Hamiltonian of the projected band is the sum of the band structure and the interaction $\bar H = H_0 + \bar H_\text{int}$. All momentum sums are over the Brillouin zone. We focus on systems sizes with $L_x = 6$ and $L_y \in \{4, 5\}$, allowing for the relevant filling fractions $\nu \in \{-1/3, -2/3, -1/2\}$. Since the system has a discrete translation symmetry, the total lattice momentum is conserved, and we diagonalize the Hamiltonian for each total momentum sector. In Fig.~\ref{fig:ED}, we show the lowest eigenvalues for each momentum sector for $\nu = -1/3$ and $\nu=-2/3$. 
In practice, to achieve a hole filling of $\nu < 0$, we fill the projected Chern band with $(1+\nu)L_xL_y$ electrons.
As expected, the ground state is threefold degenerate, and a clear gap to the first excited states is visible. Under the insertion of flux, which can be achieved by shifting the single-particle momentum $k_i \rightarrow k_i + \phi/L_i$, the three ground states flow into each other, while the gap to the excited states remains for all $\phi$. The three ground states return to their original configuration only after inserting three flux quanta.
For the numerical computation of the spectral function, we focus on a system size of $L_x\times L_y = 6 \times 4$, to allow for sufficient energy resolution. The single-particle spectral function is the imaginary part of the propagator
\begin{equation}
    \Gamma(\kk, \omega) = - \frac{1}{\pi} \braket{\psi_\kk}{x_\kk(\omega)},
\end{equation}
with $\ket{\psi_\kk} = \gamma_\kk \ket{0}$ for the hole spectral function and $\ket{\psi_\kk} = \gamma^\dagger_\kk\ket{0}$ for the particle spectral function, where $\ket{0}$ is the ground state. We obtain the state $\ket{x_\kk(\omega)}$ by iteratively solving the equation 
\begin{equation}
    (\omega + i \eta + E_0 - \Bar{H}) \ket{x_\kk(\omega)} = \ket{\psi_\kk},
\end{equation}
for each momentum $\kk$ and energy $\omega$.
Here, $E_0$ is the ground-state energy of the projected Hamiltonian $\Bar{H}$, and a small $\eta=1$~meV is used for artificial broadening.

\bibliography{bibfile}

\end{document}

%% file: newcommands.tex
\newcommand\numberthis{\addtocounter{equation}{1}\tag{\theequation}}

\newcommand{\eps}{\varepsilon}

\newcommand{\kk}{\mathbf{k}}

\newcommand{\pbold}{\mathbf{p}}

\newcommand{\qb}{\mathbf{q}}

\newcommand{\tfr}{t}
\newcommand{\bfr}{b}